\documentclass[12pt]{article}
\usepackage[dvips]{graphics,color,epsfig}
\newcommand{\klpnn}{$K^0_L \to \pi^0 \nu \bar\nu$}
\newcommand{\kpnn}{$K^+ \to \pi^+ \nu \bar\nu$}

\newcommand{\kpitwo}{$K_{\pi2}^0$}

\title{$K_L\rightarrow \pi^0\nu\bar{\nu}$ at the AGS (E926)
\thanks{talk presented at the ``International Workshop on CP Violation
in K'', 18-19 December 1998, KEK-Tanashi, Tokyo}
}
 
\author{Akira Konaka\thanks{For the E926 collaboration
(BNL, INR(Moscow), Kyoto, U.N.M., U.B.C., Jefferson Lab.,
TRIUMF, V.P.I., Yale),
Email address: konaka@sitka.triumf.ca}
\\
{\it TRIUMF}\\
{\it 4004 Wesbrook Mall, Vancouver, B.C. V6T2A3 CANADA}
}
\date{}

\begin{document}
\maketitle

\begin{abstract}
A proposed experiment to measure 
$K_L\rightarrow \pi^0\nu\bar{\nu}$ at the AGS (E926)
is described.
It adopts a unique but general strategy of rare decay measurements
with two independent rejection criteria,
which allows reliable measurements of background levels.
The method employs hermetic veto and full kinematic
reconstruction using kaon time of flight and kinematic
reconstruction of $\pi^0$.
Backgrounds are suppressed to a level well below an
anticipated signal in the range $3 \pm2 \times 10^{-11}$.
\end{abstract}
\newpage

\section{Introduction}

  \subsection{Physics Motivation}

As discussed by the previous speakers,
the rare decay $K_L\to\pi^0\nu\bar\nu$ is unique among
potential SM observables; it is dominated by direct CP violation
\cite{LIT89} and is entirely
governed by short-distance physics involving the top quark.
As a consequence of unprecedented theoretical precision and
anticipated experimental accessibility, a measurement of \klpnn~ can
unambiguously test the SM origin of CP violation and ultimately yield
the most accurate determination of the CKM CP violating phase $\eta$.
This rare decay mode therefore provides an exceptional and unique
opportunity for making significant progress in our understanding of
flavor-dynamics and CP violation.  It is competitive with and
complementary to future measurements in the $B$ meson system.  Absence
of \klpnn~ within the expected range of about $(3\pm2)\cdot 10^{-11}$
or a conflict with other CKM determinations would certainly indicate
new physics.

  \subsection{A strategy of the rare decay measurement}

In this section, a unique but general strategy for
rare decay measurements
which was found to be essential in observing 
\kpnn~ in E787, is described.
The main challenge in such a very rare decay search is to control
the systematic error in estimating the level of backgrounds.
We have to make sure that there are no very rare software/hardware
problems or unexpected rare background mechanisms that
are not taken into account in estimating the background.
In general, these potential problems are not well represented by 
Monte Carlo simulations, and one has to go back to the
data to systematically study them.
E787 analysis uses a tool called a ``bifurcated analysis''
along with a method to protect against bias due to small statistics.
As will become clear, it is essential to have
{\it two independent rejection criteria (or cuts)} for each 
background processes.

  \subsubsection{Bifurcated analysis}

\begin{figure}[htpb]
    \begin{minipage}{0.45\linewidth}
    \large
    photon veto reversed\\
    \vspace{-0.5cm}
    $\times \sim$ 50 enhancement\\
    \centerline{\psfig{figure=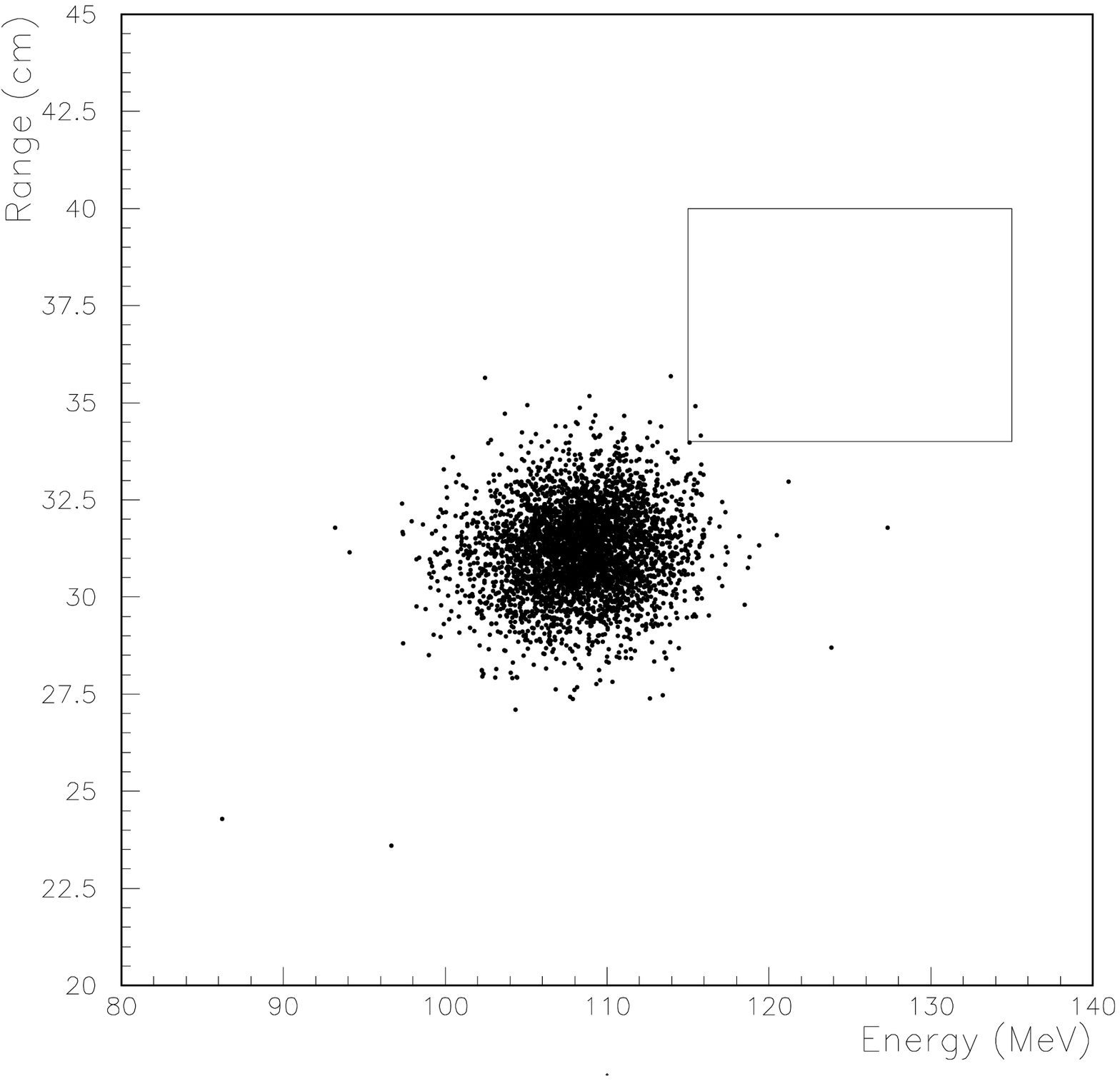,height=2.5in,width=2.8in}}
    \end{minipage}\hfill
    \begin{minipage}{0.45\linewidth}
    \large
    Selecting $K_{\pi 2}$ peak\\
    $\times \sim$ 100000 enhancement\\
    \vspace{-1.4cm}
    \flushleft{\psfig{figure=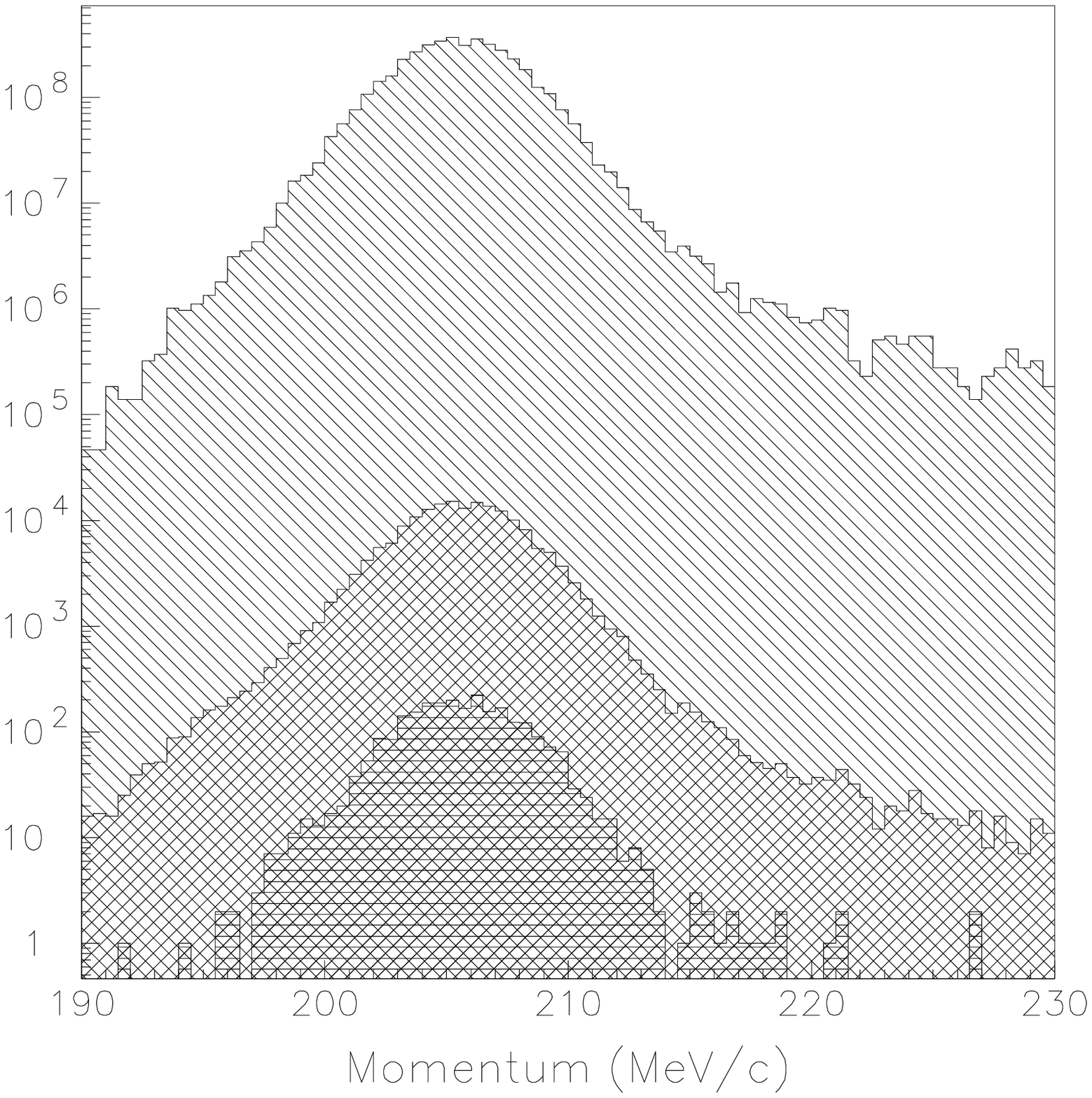,height=2.5in,width=2.8in}}
    \end{minipage}\hfill
    \caption{An example of a bifurcated analysis in the E787
       $K^+\rightarrow \pi^+\pi^0$ background analysis.}
    \label{f:bif}
    \end{figure}

The bifurcated analysis of \kpnn~ in the E787 experiment
consists of two built-in independent selection criteria
which are used to {\it measure} each background.
Figure~\ref{f:bif} shows $K^+\rightarrow \pi^+\pi^0$
background measurement as an example.
In this example,
the photon veto, which rejects $\pi^0\rightarrow \gamma\gamma$ decays,
and the charged particle
kinematics (range, energy and momentum), which rejects
the monochromatic $\pi^+$, are taken as the two independent cuts.
Because these two cuts address different particles in the
final state, they are independent in principle.
The scatter plot on the left in Figure~\ref{f:bif} shows the energy versus
range distribution when the photon veto cut is loosened. With this
sample enhanced by a factor of 50, the rejection of the kinematic
cut can be measured.
The histogram on the right shows the momentum distributions before (the
largest histogram) and after (the smallest histogram) the photon veto.
The momentum peak is suppressed by a factor of $10^{6}$.
It should be pointed out that the momentum line shape did not
change after this suppression, providing confidence
that the two criteria are independent.
All the kinematic variables used in the analysis
are tested this way to see if there are any correlation
with photon veto left.

\begin{figure}[htpb]
    \centerline{\psfig{figure=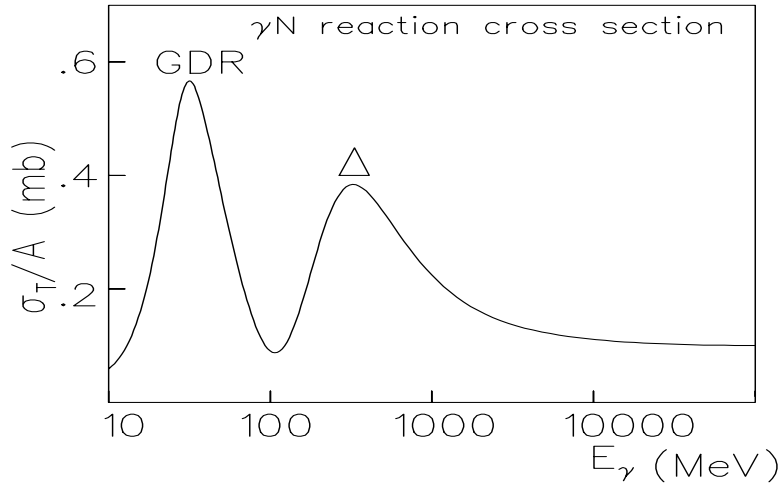,height=2.5in,width=4.0in}}
    \caption{Photonuclear reaction cross section as a function
     of photon energy. The giant
    dipole resonance of nuclei is at around 20MeV, and the
    delta resonance of the nucleon is at around 300MeV.
    The cross section converges to an asymptotic value at high energy.}
    \label{f:pn}
    \end{figure}

The bifurcated analysis is a powerful way
of measuring the background levels and their systematic uncertainties
due to correlations.
Because one can enhance the background by a large number,
it also allows us to examine rare mechanisms and
to develop detailed analysis cuts to reject them.
    \begin{figure}[htpb]
    \vspace{-1.5cm}
    \centerline{\psfig{figure=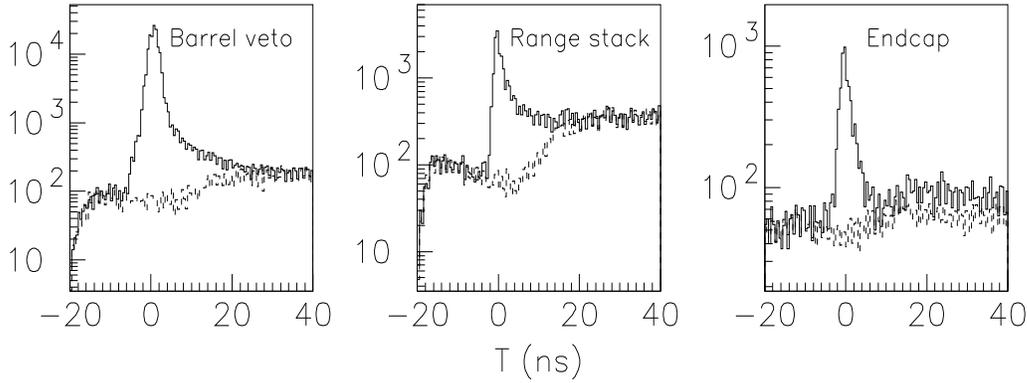,height=2.7in,width=6.0in}}
    \vspace{-0.5cm}
    \caption{The time difference between the
    charged track and extra junk energies for the $K^+\rightarrow \pi^+\pi^0$
    kinematic peak events.}
    \label{f:pvtime}
    \end{figure}
Figure~\ref{f:pvtime} shows time differences between 
the charged track and extra energy
for the $K^+\rightarrow \pi^+\pi^0$ kinematic peak events
in the 3 types of photon veto systems.
The barrel veto
is a lead/plastic scintillator sandwich,
the range stack is a stack of plastic scintillators,
and the endcap is made of pure CsI crystals.
The solid lines show the data and the dashed lines show
measured accidental rate from $K^+\rightarrow \mu^+\nu$ events.
For the barrel veto and range stack,
there are non-Gaussian tails at larger times,
which are understood as neutrons produced by photonuclear
reactions (Figure~\ref{f:pn}).
Neutrons from photonuclear reactions leave late energy through n-p
scattering in the plastic scintillation counters
in the range stack and barrel veto.
Using these data, photon veto cut positions in time versus energy space
were optimized.

  \subsubsection{Protection against bias due to small statistics}

  From time to time, we see experimental evidence of a several
standard deviation ($\sigma$)
effect appear and then disappear as more statistics are accumulated. 
This can be understood considering only statistical effects,
once we recognize the fact that one measurement out of a large 
number of experiments being performed 
should show such a large statistical fluctuation.
The same thing would happen even in 
a single peak search experiment, such as Higgs search:
Suppose we look for a peak in a distribution
which spans a range 300 times larger than the width of the expected
signal peak, we would expect on average one 3$\sigma$ peak,
whose probability is 0.27\% ($300\times 0.27\% \sim 1$).
Confusion is caused when the number of trials,
e.g. 300 of peak search regions, is not taken into account.
This is often referred to as analysis bias.

The same bias shows up in rare decay analyses.
Suppose we are left with one candidate event
and found that one signal selection criterion, which is
99\% efficient, removes it. We might conclude that
the event is a background. However, if 100 such
99\% efficient criteria are tried
before finding the one criterion that removes the event,
it is likely we would succeed in finding one even for
a signal event ($100\times 1\% = 1$).
Again, what is important is to take the number of trials into
account. In reality, we often look at an event
display and various quantities to make a physics
judgment on the event, in which procedure 
it is very difficult to quantify how many trials are
performed.
One way, which may be the only way, to protect against this analysis bias
is to avoid such procedures in which one cannot quantify
the number of trials.

  There were three steps taken by the rare decay analysis,
each of which required protection against potential bias:
\begin{itemize}

\item Candidate event selection ({\bf Blind analysis})\\
      If we examine individual candidate events and
develop selection cuts, we introduce analysis bias
as discussed above. 
The way to avoid this is to develop cuts without
examining the candidate events (``blind analysis'').
The bifurcated analyses provide large data samples 
which are good representations of background.
These samples allow us to develop
analysis cuts without examining candidate events.

\item Background measurement ({\bf Training/Test samples})\\
      If there are no events left (upper limit only) at the end,
the blind analysis might be sufficient.
However, in order to observe the signal,
one needs to avoid bias in the background measurement,
to prove if the remaining events are signal
or background.
Again, when we examine each of the remaining
background events, making many trials to find
effective cuts and adopt only the ones that
eliminate the remnant,
we introduce a bias, namely we underestimate
the background level.
Because bifurcated analysis provides large background
samples, we can afford to split them into two and use one for
developing cuts (training sample) and the other for measuring
the resulting background levels (test sample)
to avoid the potential bias.

\item Detailed examination of the candidates
             ({\bf Signal/background likelihood})\\
      It is impossible to fully prove that there is nothing
missed at all in the analysis at the end. In the case of
bifurcated analysis, there may be a hidden correlation
between two ``independent'' cuts which does not show up
in the correlation tests as described in the
previous section. We would eventually be forced to examine
the candidate event to see if there is anything
wrong with it.
As discussed above, we could always find a distribution
in which the candidate event is near the edge
    \begin{figure}[htpb]
    \centerline{\psfig{figure=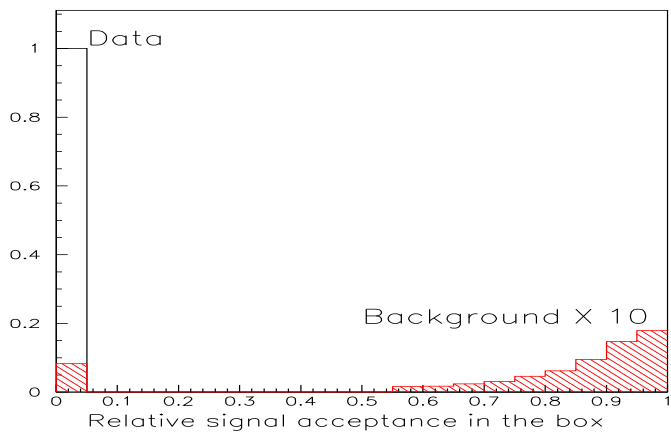,height=2.5in,width=4.0in}}
    \caption{Signal likelihood function of E787 in the signal box.
    The shaded histogram shows the expected background distribution
    scaled up by a factor of 10. The candidate event is found to be
    in the cleanest region with a factor of 10 more rejection.}
    \label{f:prob_in}
    \end{figure}
even for the real signal event,
if we look at enough quantities (enough trials).
In order to examine the candidate event in an unbiased way,
a signal likelihood function was prepared
beforehand in the E787 analysis (Figure~\ref{f:prob_in}).
The large background sample from bifurcated analysis
provided enough statistics even inside the signal region. 
There was more confidence in the signal,
because the event was in the cleanest region in the
signal box with a factor of 10 more background rejection.
It should be pointed out that this last step was feasible only
because the experiment had extra background rejection capability
(redundancy) beyond the final cut position.
\end{itemize}

\section{Basic concept of the experiment}

\begin{figure}[htpb]
    \centerline{\psfig{figure=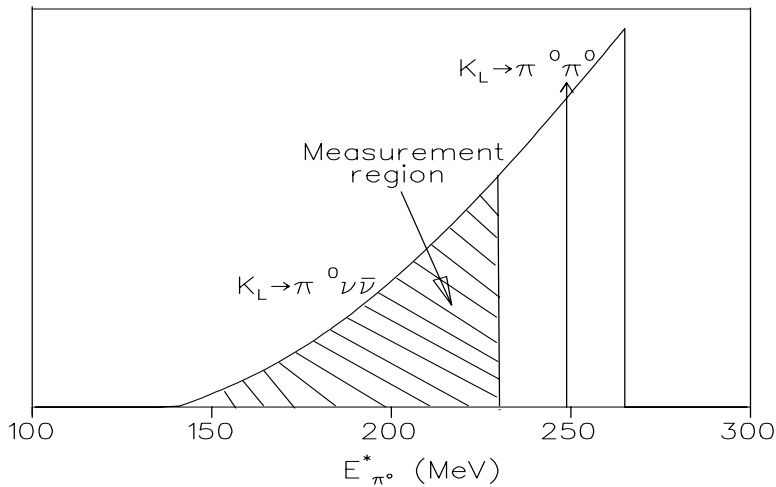,height=2.5in,width=4.0in}}
    \caption{Energy spectrum of $\pi^0$ in the $K_L$ center of mass
         frame.}
    \label{f:ecm}
    \end{figure}

Figure~\ref{f:ecm} shows the $\pi^0$ energy distribution in the
$K_L$ center of mass frame ($E_{\pi^0}^*$)
for the $K_L\rightarrow \pi^0\nu\bar{\nu}$ and
$K_L\rightarrow \pi^0\pi^0$ (\kpitwo~) decays.
The main background comes from the CP violating decay
$K_L\rightarrow \pi^0\pi^0$ 
where 2 of the photons are missed.
Even with the branching ratio of $9\times 10^{-4}$
and photon detection inefficiency
of $10^{-4}$ per photon (the value that is achieved
for higher energy photons in E787),
the background appears at a single event sensitivity
of $(9\times 10^{-4})\cdot(10^{-4})^2\cdot_4C_2=5.4\times 10^{-11}$.
This is larger than the central value of the Standard Model prediction
of $3\times 10^{-11}$.
As we have seen in the previous section,
it is very difficult to measure the background with only
a photon veto cut.
Thus, to increase the
probability that the source of observed ``$\pi^0$ plus nothing'' events is
truly the $\pi^0 \nu \bar{\nu}$ mode another handle
using {\it kinematics} is needed.
By measuring the decaying K$^0_L$ momentum using time-of-flight,
one can transform the $\pi^0$ momentum back to the $K_L$
center of mass system on an event-by-event basis, and
reject the monochromatic \kpitwo~ peak.
With these two independent cuts of kinematics and photon veto, 
we can get sufficient background rejection and make a
reliable measurement of the background.

    \begin{figure}[htpb]
%    \centerline{\psfig{figure=design2.ps,height=0.7in}}
    \begin{minipage}{0.38\linewidth}
    \centerline{\psfig{figure=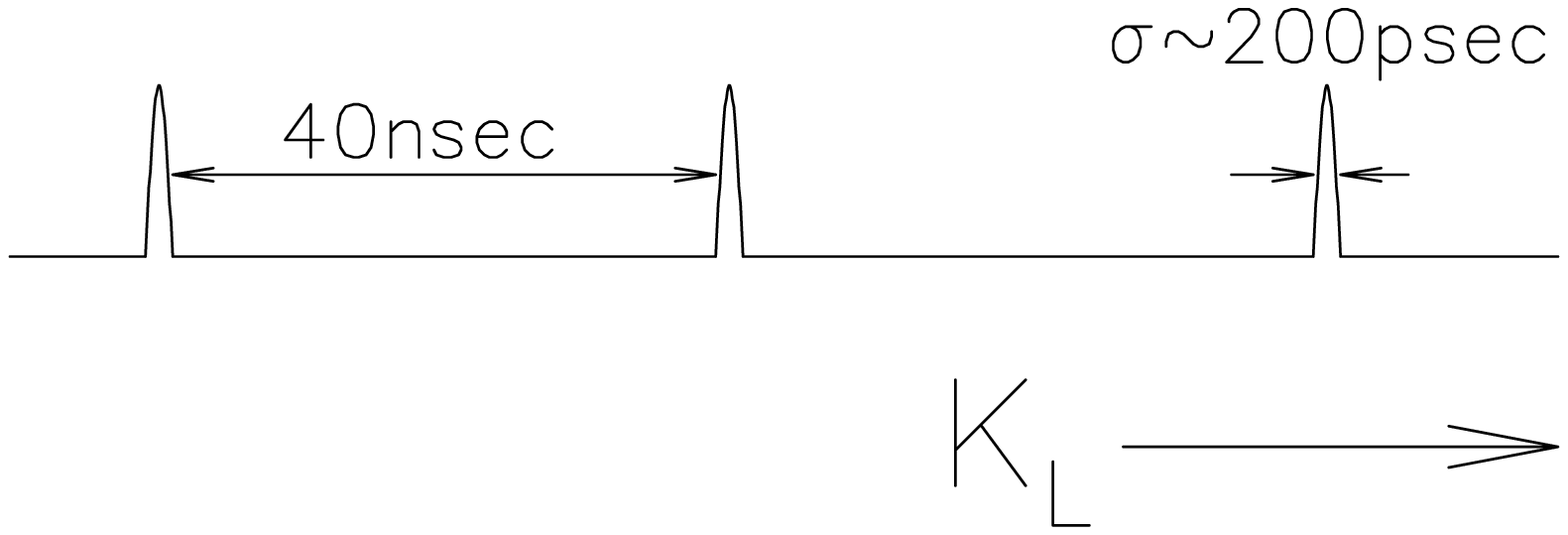,height=1.5in,width=3in}}
    \vspace{1in}
    \end{minipage}\hfill
    \begin{minipage}{0.52\linewidth}
    \centerline{\psfig{figure=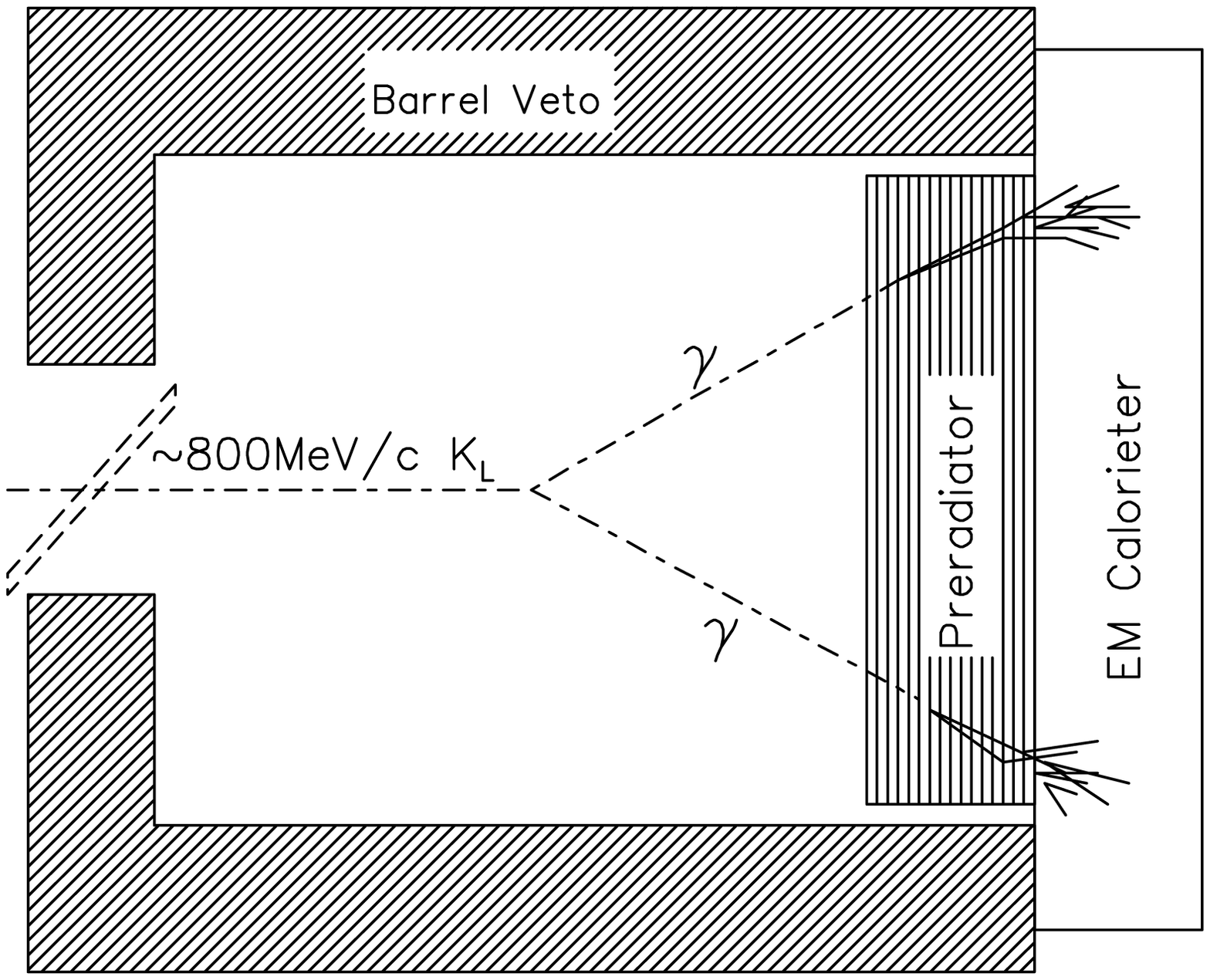,height=3in,width=4in}}
    \end{minipage}\hfill
    \caption{A schematic view of the detector.}
    \label{f:schem}
    \end{figure}

Figure~\ref{f:schem} shows a schematic view of the detector.
A bunched $K_L$ beam produced by the pulsed AGS proton beam 
decays in the decay region which is hermetically surrounded by
photon detectors. The directions of the photons are
measured by a preradiator and the total energy is measured
by an electromagnetic calorimeter. In the following
sections, more detailed descriptions of the kinematic criteria
and photon veto are given.

  \subsection{Full kinematic reconstruction and
         particle identification}

Since the 3-body $\pi^0$ spectrum is the only observable in \klpnn,
the most effective strategy for a clear measurement
is to fully reconstruct the $\pi^0$ in the
$K_L$ center of mass frame.
This can only be efficiently and
unambiguously accomplished by completely measuring the kinematics of
the decay photons including time, position, angle and energy and by
determining the $K_L$ momentum by time-of-flight from the production
target.
This method results in positive identification of the signal
and effective suppression of the background
since, for example, the two-body \kpitwo~ decay  identifies
itself by the unique momentum of the $\pi^0$ when viewed in the $K_L$
rest frame.  Once the $K_L$ momentum
is known, a large fraction of the
\klpnn~ phase space is available for detection and all the major
sources of background become manageable.

\begin{figure}[htpb]
    \centerline{\psfig{figure=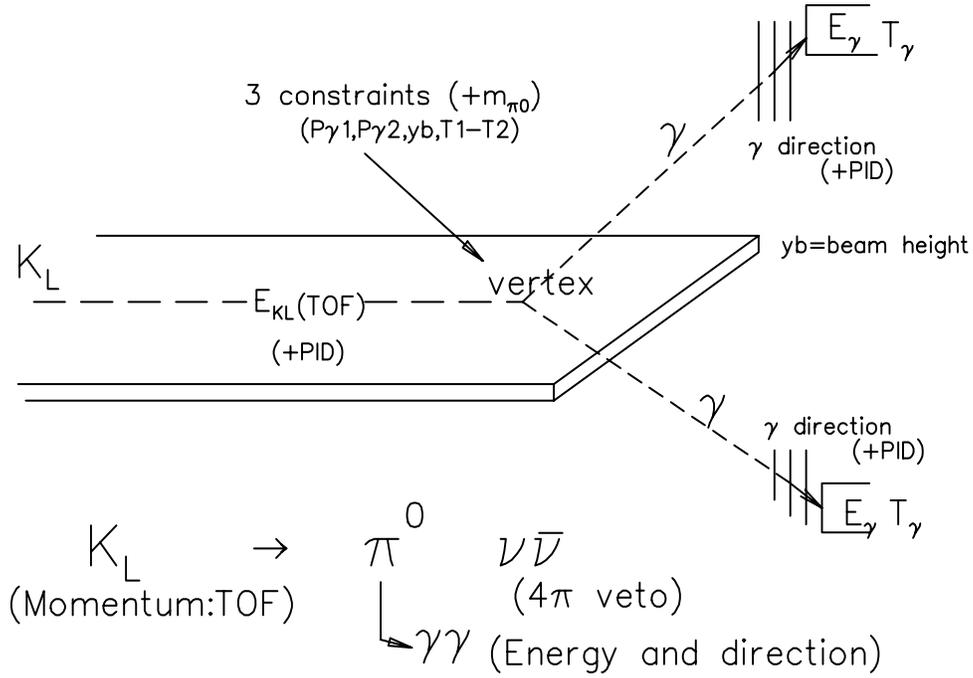,height=3.5in,width=5.0in}}
    \caption{Quantities measured in the experiment.}
    \label{f:full_kin}
    \end{figure}

    Figure~\ref{f:full_kin} shows the measured quantities
  in the experiment. The energies and times of the photons
  ($E_\gamma,T_\gamma$) are measured by the calorimeter
  and the directions are measured by the preradiator.
  The preradiator also accomplishes particle identification:
  The first layer of scintillator identifies whether
  it is a charged or neutral particle, and dE/dx and
  the track pattern discriminates
  photons from $K_L$ and neutrons.

  The $K_L$ beam is collimated into a plane (slit
  beam). The decay vertex is measured by extrapolating
  the photon direction into the beam plane. The vertex
  can be obtained from one photon direction alone;
  the direction of the other photon, the time difference
  between the two photons, which corresponds to the path length
  difference between the two photons, and the $\pi^0$ mass
  work as extra constraints. These extra constraints are
  effective in removing non-Gaussian tails, particularly in
  the angular
  measurements by the preradiator.
  The $K_L$ energy is measured by time of flight
  using on RF bunched beam and direction is given
  by connecting the production target position and
  the decay vertex.

  \subsection{Photon veto}

   E787 has achieved a $\pi^0$ detection inefficiency of $10^{-6}$ at
   photon energies of 20--225 MeV using a lead/scintillator
   detector.
   The inefficiency for the lowest
   energy photons is $\sim 10^{-2}$ and appears to be mainly limited
   by sampling fluctuations.  The inefficiency for higher energy
   photons is $\sim 10^{-4}$ and appears to be limited by sampling
   fluctuations, shower escape and photonuclear reactions which may be
   contributing at comparable levels.  Although the E787 group has
   attempted to study the origin of the residual inefficiency through
   measurements and simulations, considerable uncertainty remains.
   Thus, assuring the achievement of substantially higher photon
   detection efficiencies, for instance, at higher photon energies
   than the existing measurements, may be extremely difficult to
   establish reliably without actually performing the relevant
   measurements. The uncertainties are even larger for the region
   below 20 MeV photon energy which will be preferentially populated by
   $K^0_{\pi 2}$ background events involving higher energy $\pi^0$'s than
   observed in E787.  It is for these reasons that, although
   advancements in detection efficiency may be possible to achieve, we
   will only rely on small extrapolations from the E787 measurements
   in predicting the level of $\pi^0$ inefficiency.

   The goal in the \klpnn~ experiment is to have $\pi^0$ detection
inefficiency approaching $10^{-8}$. This is feasible since both
photons are generally in the higher energy range of the E787
measurements where single photon detection inefficiencies of $10^{-4}$
have been measured.  Since we also have kinematic handles available,
we can suppress those kinematic configurations of \kpitwo~ events with
low energy missing photons and reasonably expect to achieve the goal.

\begin{figure}[htpb]
    \centerline{\psfig{figure=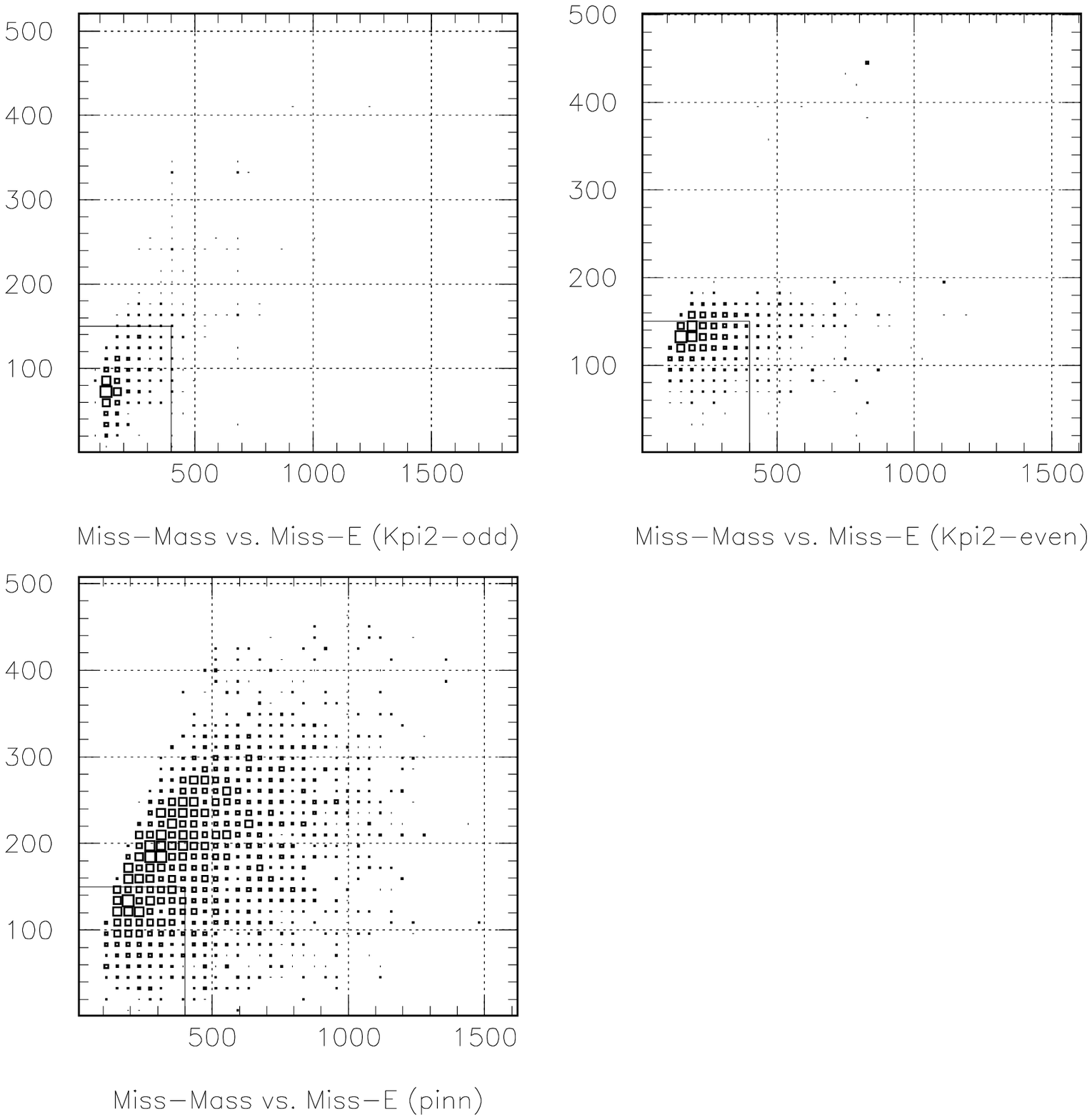,height=4.in,width=4in}}
    \caption{Missing mass vs. missing energy distribution of
     photons for \kpitwo~odd, \kpitwo~even and \klpnn~ events.}
    \label{f:misscut}
    \end{figure}

   Because the momentum of the $K_L$ is tagged, we can obtain the energy
   of the missing photons in \kpitwo~ events by subtracting the
   measured energies of the two observed photons from the $K_L$
   energy.  Requiring significant total missing energy (i.e.
   ($E(K_L)-E_{\gamma 1}-E_{\gamma 2}$) as is generally the case for
   \klpnn~ events suppresses most potential background events that
   contain lower energy missing photons (where the inefficiency is
   greatest). However, in unusual cases when one of the missing
   photons has very high energy and one has very low energy an
   additional cut on missing mass (i.e. $\sqrt{(E(K_L)-E_{\gamma
   1}-E_{\gamma 2})^2- ({\bf{P}}(K_L)-{\bf{P}}_{\gamma
   1}-{\bf{P}}_{\gamma 2})^2}$) is effective. Because the missing mass
   in \kpitwo~ events is proportional to $\sqrt{E_{miss1}*E_{miss2}}$,
   where $E_{miss}$ is the energy of a missing photon, the missing
   mass also becomes small for the asymmetric energy sharing case.
   Figure~\ref{f:misscut} shows the missing mass vs. missing energy
   distribution of photons for \kpitwo~
and  \klpnn~ events.  After removing the low missing mass and
   low missing energy region, we can suppress the low energy photons
   to achieve $10^{-8}$ detection inefficiency for the two missing
   photons in $K^0_{\pi 2}$ events.

\begin{figure}[htpb]
    \centerline{\psfig{figure=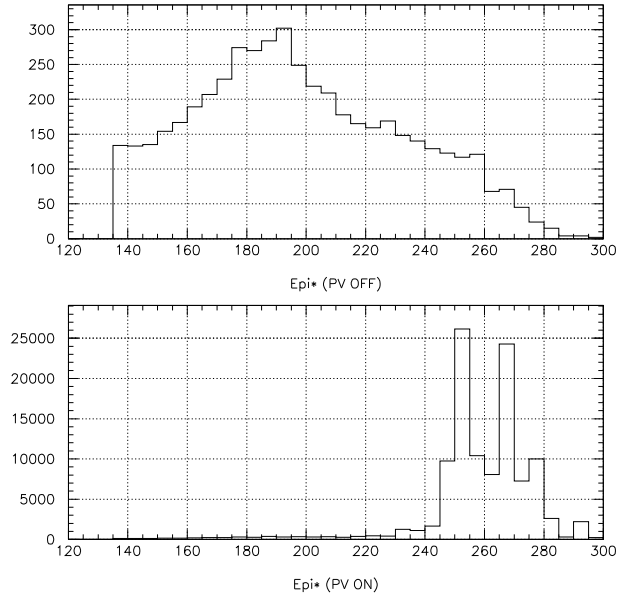,height=3.5in,width=3.5in}}
    \caption{$\pi^0$ energy distribution in the $K_L$ center of mass
     system ($E_{\pi^0}^*$)
     for \kpitwo~odd events before (top) and after (bottom)
     photon veto cuts. The vertical scale of the bottom figure
     is in the unit of $10^{-8}$ events.}
    \label{f:pv_on_off}
    \end{figure}

The effect of small missing mass can be seen more directly by
comparing $E_{\pi^0}^*$ distributions
before and after the photon veto cut
for \kpitwo~odd background events
where one photon from each $\pi^0$ is missed
(Figure~\ref{f:pv_on_off}).
A dramatic peak above the $E_{\pi^0}^*$=230MeV
after photon veto cut directly corresponds
to the small missing mass
\footnote{There is a one-to-one correspondence
between $E_{\pi^0}^*$ and missing mass:
$E_{\pi^0}^*=\frac{m_K^2+m_{\pi^0}^2-m_{miss}^2}{2m_K}$.
For small
missing mass, $E_{\pi^0}^*$ is large.}.
This is one of the main reasons why the phase space
below \kpitwo~ peak ($E_{\pi^0}^*$=249MeV) is used
for the signal search region.

\section{Experimental design}

 \subsection{An overview}

\begin{figure}[h]
  \centerline{\psfig{figure=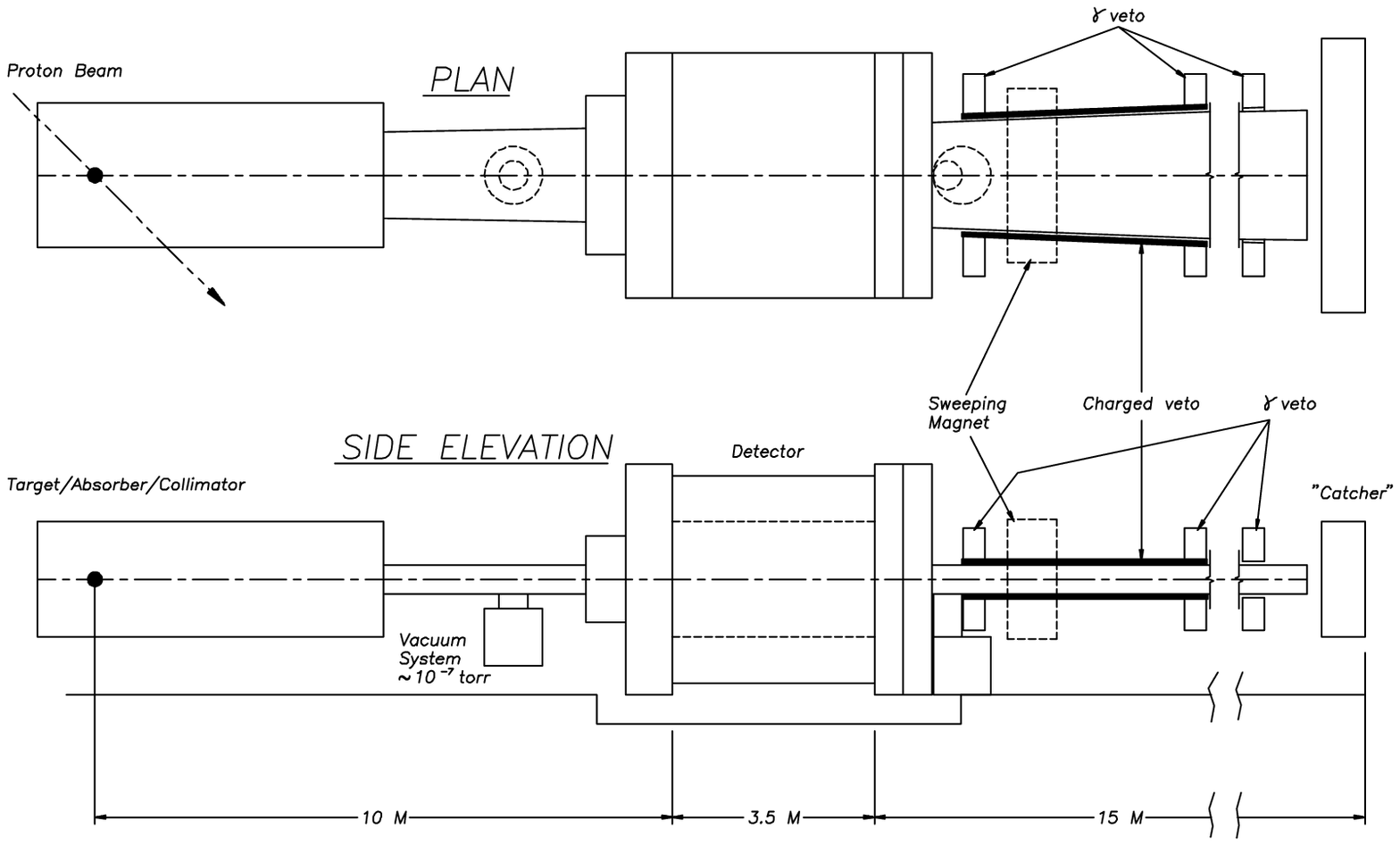,height=3.5in}}
  \caption{\label{layout} Layout of the \klpnn~ experiment.}
    \end{figure}

Figure ~\ref{layout} shows a layout of the proposed 
experiment.
A 500 $\mu sr$ solid angle neutral beam is extracted at
45$^o$ to produce a ``soft'' $K_L$~ spectrum
peaked at 0.65 GeV/c. Kaons in the range from about 0.5
GeV/c to 1.3 GeV/c will be used.  The vertical acceptance of the beam
(0.004 r) is kept much smaller than the horizontal acceptance (0.125
r) so that effective collimation can be used to severely limit
beam halos.  Downstream of the final beam collimator is a 3.5 m long
decay region which is surrounded by the main detector.  Approximately
16\% of the kaons decay yielding a decay rate of about 25 MHz.
The beam region is evacuated to a level of $10^{-7}$~Torr
to suppress neutron induced $\pi^0$ production. The decay region is
surrounded by an efficient Pb/scintillator photon veto detector.
A \v{C}erenkov detector (``catcher'') covers the beam hole
at the end of the beam line, 15 m downstream of the detector.

  \subsection{The $K_L$ beam}

In recent years the AGS has achieved new records of intensity for
synchrotrons. The present SEB peak extraction current is 
$7\times 10^{13}$ protons/pulse.
Coupled with a  high current  micro-bunched beam, good duty
factor and extended availability during the RHIC era,\footnote{ RHIC
is projected to operate for 30 to 40 weeks per year and requires
injection from the AGS for  2 hours/day. Thus,
approximately 22 hours/day are available for AGS proton operation.} the
AGS is the ideal accelerator site for rare neutral kaon
decay experiments employing time-of-flight.
    \begin{figure}[htpb]
    \begin{minipage}{0.36\linewidth}
    \centerline{\psfig{figure=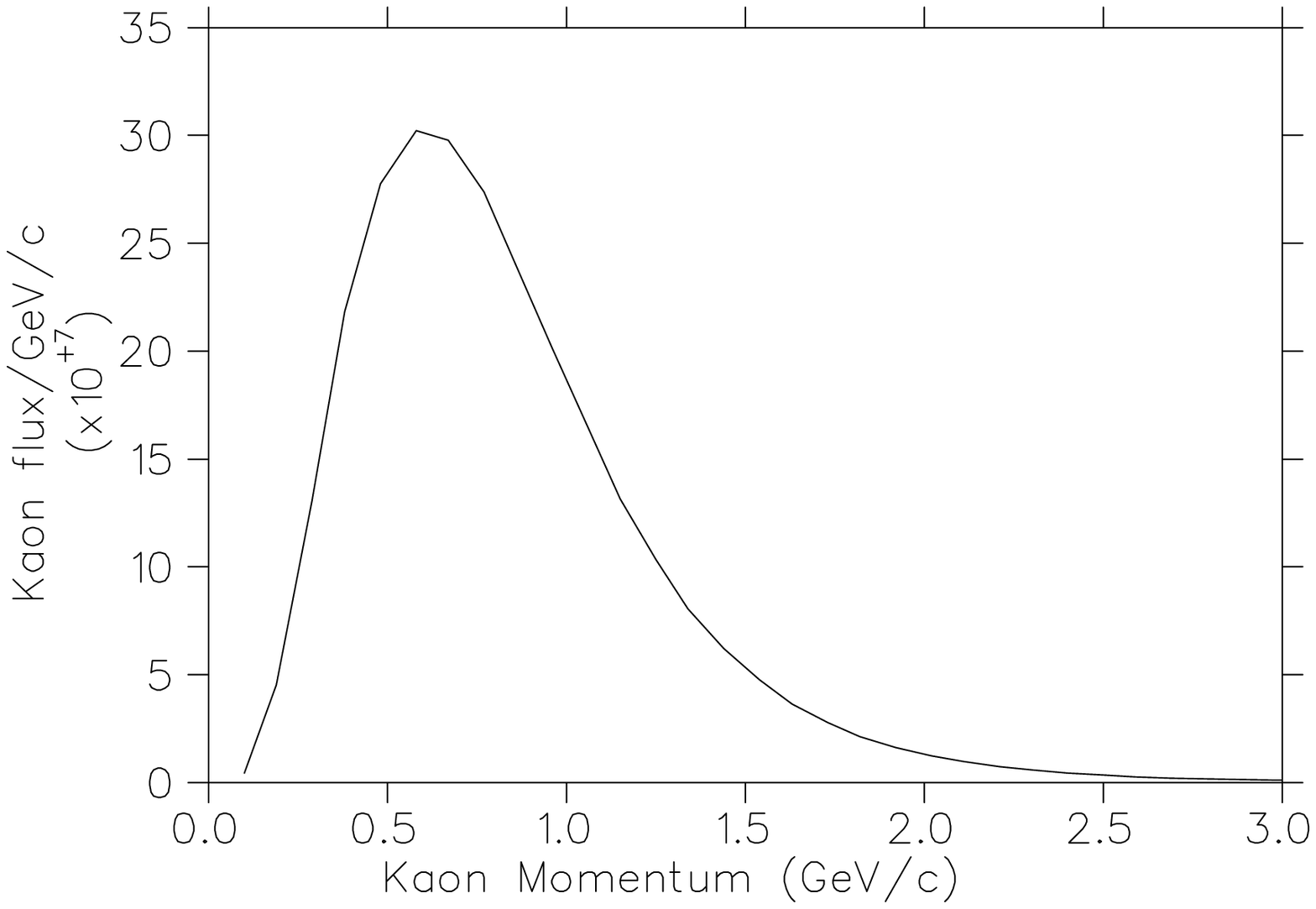,height=2.5in,width=3.2in}}
    \end{minipage}\hfill
    \begin{minipage}{0.5\linewidth}
    \vspace{-.4in}
    \flushleft{\psfig{figure=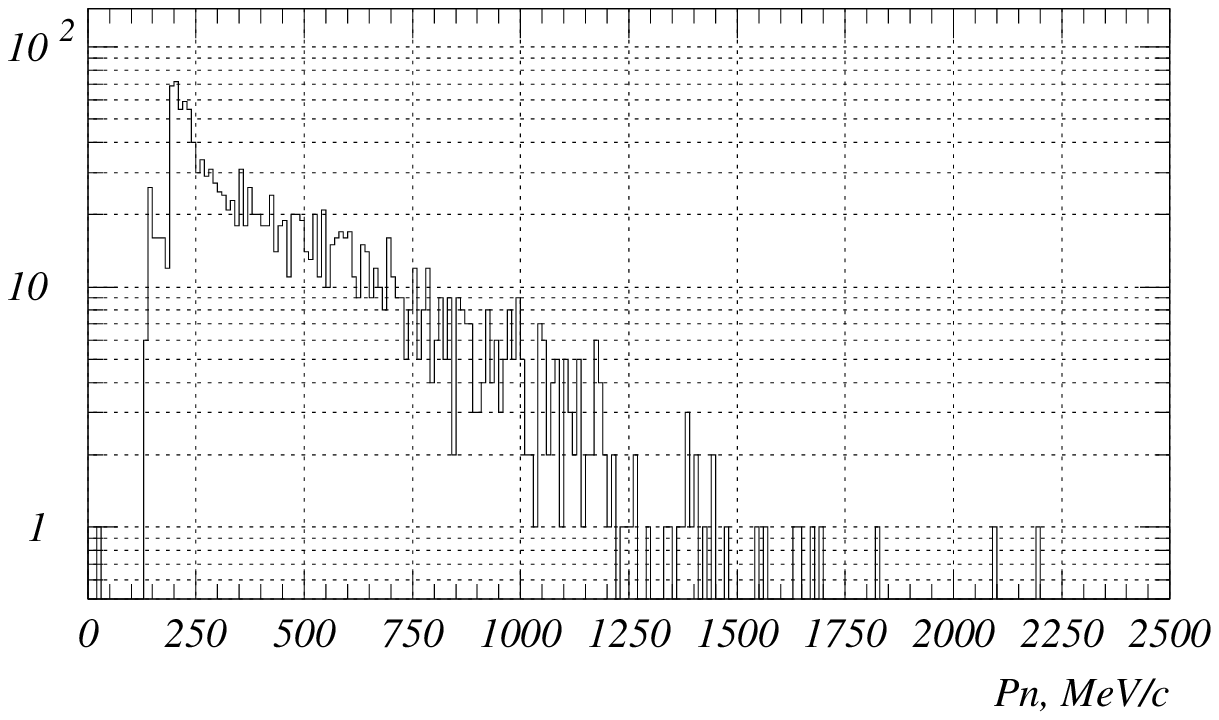,height=3.5in,width=3.5in}}
    \end{minipage}\hfill
    \caption{Momentum distribution of $K_L$ and neutron fluxes}
    \label{f:kflux}
    \end{figure}
The spectrum of neutral kaons produced at 45 degrees 
shown in Figure~\ref{f:kflux} (left) was obtained using
a code based on a fit to CERN
charged particle production data\cite{cern80}.
The spectrum shape is consistent with 
large angle charged kaon production data (BNL-E802) fitted
by semi-phenomenological analytical methods\cite{e802}\cite{kapinos}.
The $K_L$ production yield is consistent within a factor of 2
between these two estimations.

Neutrons from the production target could
limit the beam intensity to be tolerated.
In particular, neutrons above the pion production threshold
can cause accidental backgrounds.
The neutron momentum spectrum for $\rm{45^\circ}$ is shown in 
Figure~\ref{f:kflux} (right).
At this large production angle,
the neutron flux is largely suppressed above 800MeV/c,
the pion production threshold of the n-p interaction.

    \begin{figure}[htpb]
    \centerline{\psfig{figure=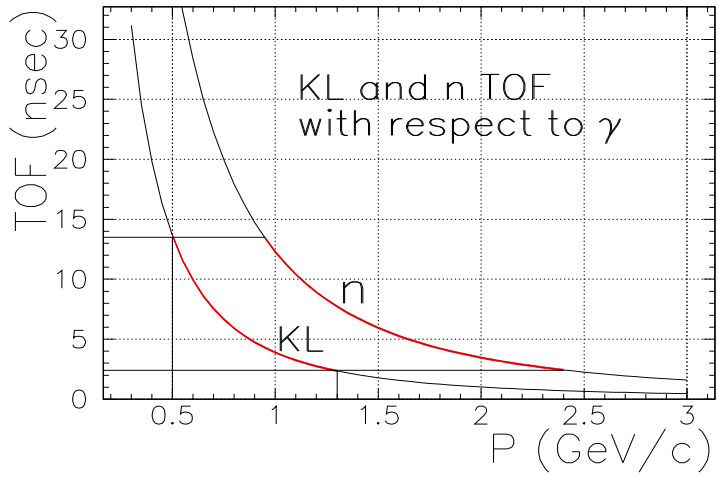,height=2.8in,width=4.0in}}
    \caption{Arrival time of $K_L$'s and neutrons with respect to
     photons at 10m downstream from the production target.}
    \label{f:tof}
    \end{figure}

The micro-structure of the beam provides further
suppression of the neutrons.
Figure~\ref{f:tof} shows the arrival time of
$K_L$s and neutrons with respect to photons
at 10m  from the production target.
Neutrons with momenta between 0.9 and 2.4 GeV/c
fall into  our arrival time of  interest (i.e.
$K_L$ with momenta between 0.5 and 1.3 GeV/c).
Within this time window,
the neutron to $K_L$ ratio is improved by a factor of 5.
Despite the fact that a low energy beam is used here,
the effective  n/$K_L$ ratio  is as good as or better than
in higher energy experiments.

  \subsection{The detector}

\begin{figure}[htpb]
    \centerline{\psfig{figure=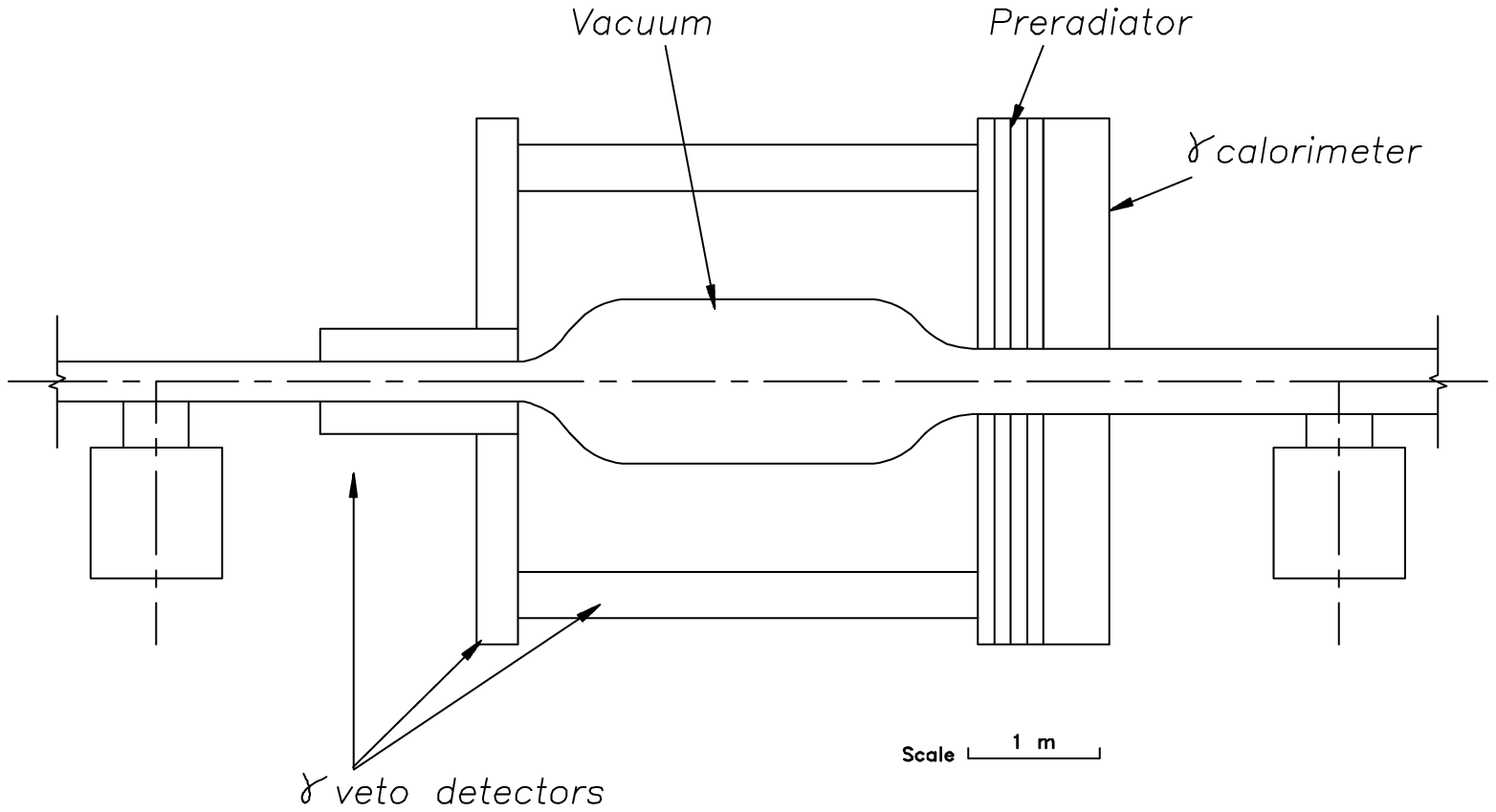,width=5.25in}}
\caption{\label{detector} \klpnn~ detector.}
\end{figure}

In the forward detection region the primary photon detector system
illustrated in Figure~\ref{detector}
consists of two sections: a fine grained preradiator in which the
photons are converted and the first $e^+/e^-$ pair is tracked, followed
by an 18 radiation length (X$_0$) calorimeter in which the remaining
energy of the photon shower is measured. The preradiator consists of
42 layers each with plastic scintillator, 0.03 X$_0$ of heavy metal
and dual coordinate drift chambers. The preradiator which has a total
effective thickness of 1.5 X$_0$ functions to measure the photon
positions and directions accurately in order to allow reconstruction
of the $K_L$ decay vertex while also contributing to the achievement
of sufficient energy resolution.
The calorimeter is constructed using thin Pb sheets formed to accept
scintillating fibers in an arrangement similar to that recently made
for the KLOE experiment at DA$\Phi$NE; in our implementation of the
calorimeter, a significantly higher visible light fraction will be
used to further improve the energy resolution.
In order to reduce the cost of the detector,
we are also considering a shashlik calorimeter, which is
made of layers of thin Pb sheet and plastic scintllator read out
by WLS fibers running perpendicular to the layers.

The barrel photon detector is made of layers of 1mm lead and
7mm extruded scintillator.
The light from the scintillator is read out by embedded wave
length shifter (WLS) fibers.
A prototype test with 4-m long WLS fibers shows
a light yield of 12.1 p.e./MeV with a timing resolution
of $\sigma=$0.85nsec \cite{yury}.
Downstream of the main
$\pi^0$ detector, a beam hole photon counter consists of
layers of lead and lucite or aerogel \v Cerenkov detectors
designed to be insensitive to neutrons.
An initial prototype test using tagged photon and neutron beams
shows promising results satisfying
the experimental requirements \cite{nomura}.

\section{Expected background levels and sensitivity}

The dominant background for \klpnn~ is the CP violating decay $K_L
\rightarrow \pi^0 \pi^0$ ($K_{\pi 2}$) with a branching ratio of $9
\times10^{-4}$.  By tagging the
$K_L$ momentum as well as determining the
energy and direction of $\gamma$s, one can reconstruct the
kinematics.  In the case of even pairing where one $\pi^0$ is missing,
a kinematic cut on the monochromatic 
$E^*_{\pi^0}$ is effective.  In the case of odd paring where
one photon from each $\pi^0$ is missed, the  $\pi^0$ mass requirement
($m_{\gamma\gamma}$) is effective.
An additional photon energy cut 
$E^*_{\pi^0}$ vs. $|E^*_{\gamma 1}-E^*_{\gamma 2}|$,
where $E^*_{\gamma 1}$ and $E^*_{\gamma 2}$ are
the energies of $\gamma$s in the $K_L$ center of mass
system,
is also very effective in further
suppressing the \kpitwo~ background.

    \begin{figure}[htpb]
    \begin{minipage}{0.45\linewidth}
    \Large
    \centerline{\psfig{figure=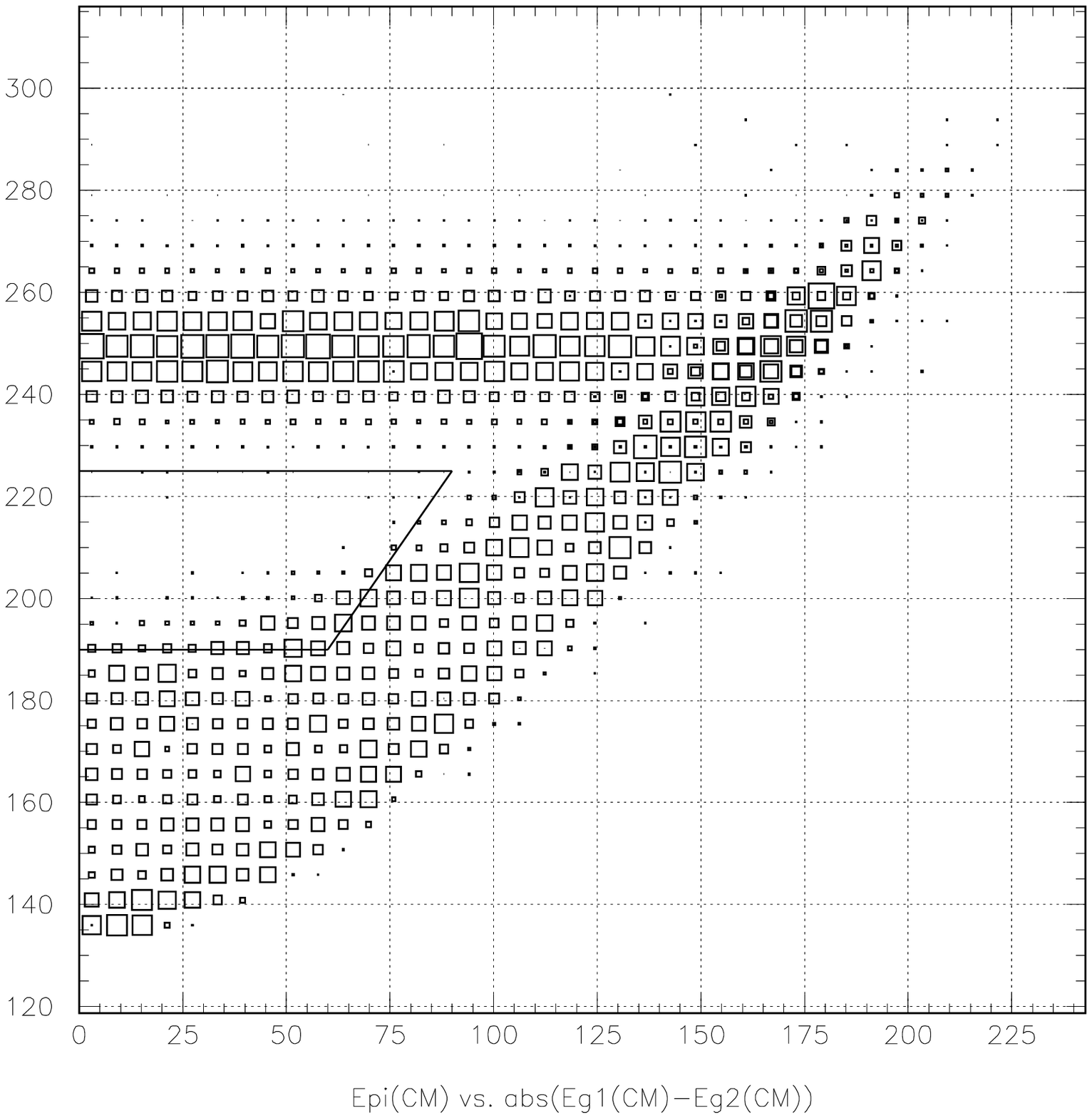,height=3.0in,width=3.0in}}
    \end{minipage}\hfill
    \begin{minipage}{0.45\linewidth}
    \Large
    \flushleft{\psfig{figure=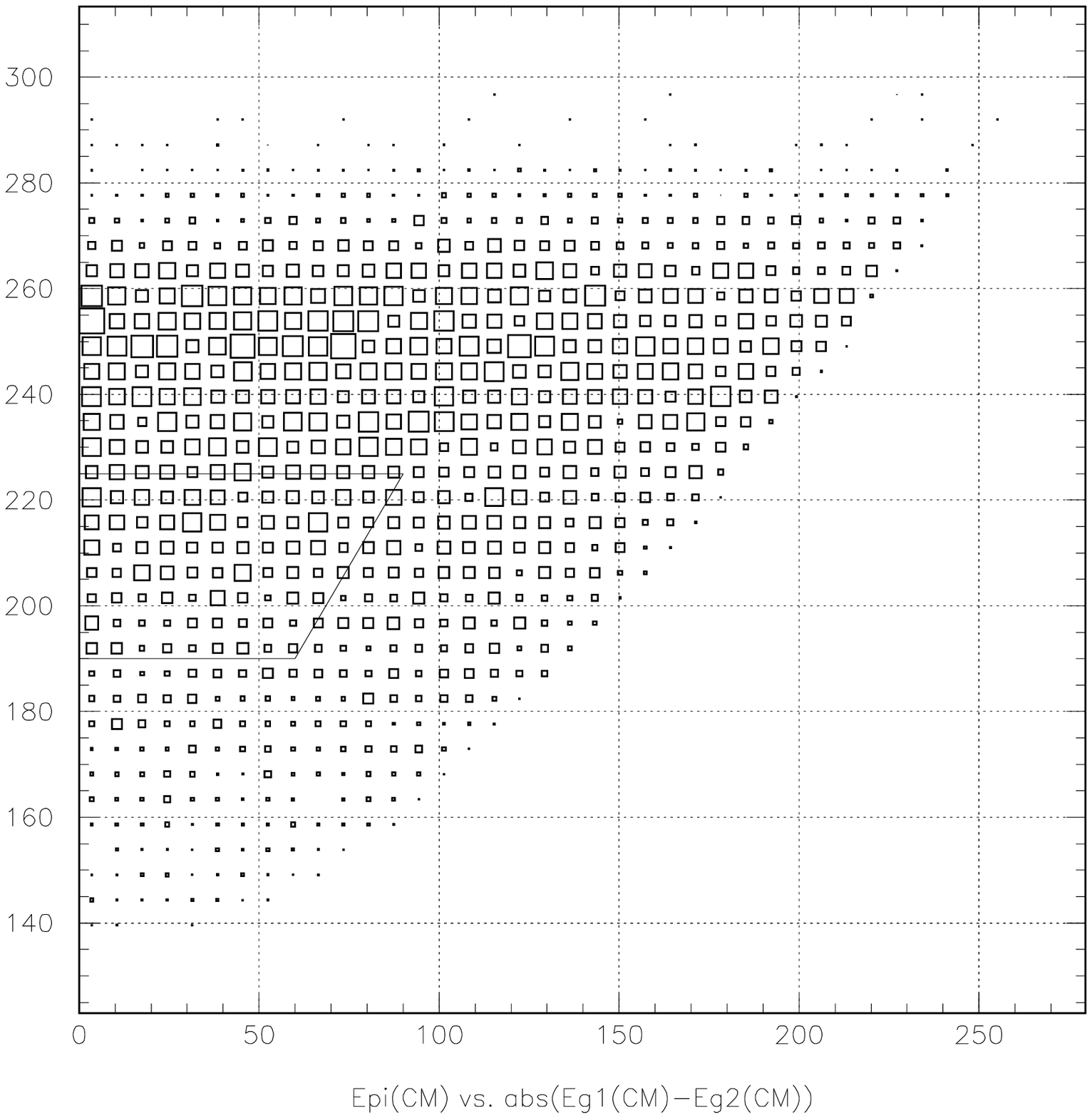,height=3.0in,width=3.0in}}
    \end{minipage}\hfill
    \caption{Distributions of
$E^*_{\pi^0}$ vs. $|E^*_{\gamma 1}-E^*_{\gamma 2}|$
after $\pi^0$ mass requirement for
the $K_L\rightarrow\pi^0\pi^0$ (left) and for the
signal (right).
The solid line encloses the signal region.}
    \label{f:ecm-vs-eg12}
    \end{figure}

Figure~\ref{f:ecm-vs-eg12} shows distributions of
$E^*_{\pi^0}$ vs. $|E^*_{\gamma 1}-E^*_{\gamma 2}|$,
after $\pi^0$ mass requirement for
the $K_L\rightarrow\pi^0\pi^0$ (left) and for the
signal (right).
The band at $E_{\pi^0}^*=$249MeV corresponds to the
even pairing background, which is suppressed by a
$E^*_{\pi^0}$ cut at 225MeV/c as discussed above.
The remaining band corresponds to the odd pairing background,
which is confined into the band
due to the constraint of 2 detected photons
to form the $\pi^0$ mass.
The solid line shows the signal region.
As can be seen in the plots,
we can obtain further background suppression at the cost
of a modest acceptance loss.
This redundancy in background rejection provides a safety
margin in the experiment as well as
a tool to positively identify the signal.
Table~\ref{acctwo} gives the estimated acceptance factors for  the
\kpitwo~ backgrounds.
\begin{table}[htbp]
\centering
\begin{tabular}{|c|r|r|r|}
\hline
Requirement & $\pi^0\nu\bar{\nu}$ & $\pi^0\pi^0$ odd & $\pi^0\pi^0$ even \\
\hline
$\gamma$ combinations &1&4&2 \\
solid angle &0.36 & 0.34 &0.37 \\
$m_{\gamma\gamma}=m_\pi$ &0.72 &0.08&0.74 \\
$E_\pi^*$ &0.38 &  0.76&0.008 \\
photon veto & & $1.3\times10^{-8}$&$4.2\times10^{-8}$ \\
Photon energy cuts & 0.35 & 0.07&0.17 \\
BR & & $9\times10^{-4}$ & $9\times10^{-4}$ \\
$\gamma$ conv. and reconst. &0.45&0.45&0.45 \\
\hline
Acceptance & $0.016$ & $3.5\times10^{-14}$ & $1.6\times10^{-14}$ \\
\hline
%Events & $73$ & 5 & 2 \\
%\hline
\end{tabular}
\caption{\label{acctwo} Acceptance for \kpitwo~ backgrounds.}
\end{table}

Other potential sources of background include
neutron production of $\pi^0$s, other $K_L$
decays like $K_{e3}$ and $K_L\to \gamma\gamma$ and $\Lambda \to n
\pi^0$ decays. Suppression of most backgrounds is accomplished by
the  hermetic high efficiency photon detector along with kinematic
constraints. 
A summary of the
background estimates is given  in
Table~\ref{t:background}.
The signal is estimated to exceed the
background by an order of
magnitude with the backgrounds dominated by  \kpitwo.
We expect that the actual background levels will be determined
reliably from the data.

\begin{table}[h]
\centering
\begin{tabular}{|c|c|c|c|} \hline
Process & Modes studied & Main  & events \\ \hline
$K_L\rightarrow \pi^0\nu\bar{\nu}$ & & & 70 \\
\hline
$K_L$ decays ($\bar{\gamma}$)
  & $\pi^0\pi^0$,$\pi^0\pi^0\pi^0$,$\pi^0\gamma\gamma$
  & $\pi^0\pi^0$ & 7 \\ 
$K_L\rightarrow \gamma\gamma$ & $\gamma\gamma$
  & $\gamma\gamma$ & 0.04 \\ 
$K_L$ decays ($\overline{chrg}$)
  & $\pi^{\pm}e^{\mp}\nu$,$\pi^{\pm}\mu^{\mp}\nu$,$\pi^{+}\pi^{-}$
  & $\pi^{-}e^{+}\nu$ & 0.01 \\ 
$K_L$ decays ($\bar{\gamma},\overline{chrg}$)
  & $\pi^{+}\pi^{-}\pi^{0}$,$\pi^{\pm}l^{\mp}\nu\gamma$,
    $\pi^{\pm}l^{\mp}\nu\pi^0$,$\pi^{+}\pi^{-}\gamma$
  & $\pi^{+}\pi^{-}\pi^{0}$ & 0.003 \\
Other decays
  & $\Lambda\rightarrow \pi^0 n,K^-\rightarrow \pi^-\pi^0,
    \Sigma^+ \rightarrow \pi^0 p$
  & $\Lambda\rightarrow \pi^0 n$ & 0.03 \\
Interactions
  & n, $K_L$, $\gamma$
  & $n\rightarrow \pi^0$ & 0.5 \\
Accidentals
  & n, $K_L$, $\gamma$
  & n, $K_L$, $\gamma$ & 0.3\\
\hline
\end{tabular}
\caption{\label{t:background}
Estimated event levels for signal and background.}
\end{table}

Table~\ref{acctwo} gives factors leading to the estimated
acceptance
of approximately 1.6\%.  It includes the solid angle, photon
conversion and reconstruction factors and phase space ($E_{\pi^0}^*$)
acceptance in addition to the photon energy cuts discussed above.
The inefficiency due to accidental spoiling of
good events is estimated to be $<10$ \% for a threshold of a few MeV
and a timing window of 2 ns.  The expected number of \klpnn~ events to
be accumulated for 8000 hours of beam
%at $5\times 10^{13}$ protons/spill
is $\sim$ 50 assuming 
the SM central value for the branching ratio of B$=3\times 10^{-11}$.

\section{Summary and prospects}
A detector system has been presented to measure the rare decay \klpnn.
The experiment is designed to allow definitive observation of
a large sample of
events with a signal that exceeds backgrounds by an order of
magnitude. 
Special features of the AGS allow provision of an intense
pulsed beam of neutral kaons suitable for time-of-flight measurements.
The two independent signal selection criteria of
photon veto and kinematics allow large background
rejection and reliable measurement of the background levels.
Significant redundancies and contingency factors are built into the
technique.
The standard model origin of CP violation will be confirmed
and the complex phase parameter $\eta$ determined to a precision of
$\leq 15 \%$ or the absence of \klpnn~ can be established at a level
inconsistent with the standard model.

%%%%%%%%%%%%%%%%%%%%%%%%%%%%%%%%%%%%%%%%%%%%%%%%%%%

\end{document}